\documentstyle[epsfig,fullpage,a4]{article}

\begin{document}
\begin{center}

{\bf Feed rotation corrections for antennas having beam waveguide mounts.}

R. Dodson, M. Rioja 

\end{center}





\section{Abstract}

We report on the development of new code to support the beam waveguide antenna mount types in AIPS, which will allow polarisation analysis
of observations made using these antennas.
Beam Wave-guide antennas in VLBI are common in communication antennas that have been repurposed (e.g. Warkworth, Yamaguchi).

The mount type affects the differential phase between the left and the
right hand circular polarisations (LHC and RHC) for different points
on the sky. 
We demonstrate that the corrections for the Warkworth beam wave guide antenna can be applied.

\section{Telescope Mounts}

Mount types are a combination of the focus position and the drive
type. In Radio Astronomy there are six more or less commonly used
focus positions, and three drive types. 

\subsection{Focus positions}

\begin{figure}[htb]
\begin{center}
\epsfig{file=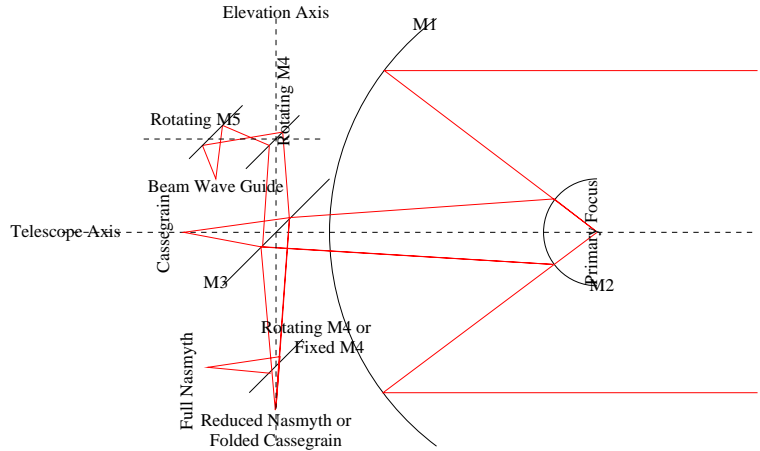,width=10cm}
\caption{Mount types, with the minimum number of mirrors (fixed or rotating with respect to the prime focus), for Prime, Cassegrain, Folded and reduced Nasmyth, Full Nasmyth and Beam Wave Guide. The first three (for an Alt-Az antenna) follow the Cassegrain feed angle rotation, the reduced and full Nasmyth introduce a dependence on $\pm$Elevation and the Beam Wave Guide introduces a dependence on $\pm$Azimuth.}
\label{fig:mounts}
\end{center}
\end{figure}

Radio telescopes use a much more limited set of focus positions compared to optical telescopes. 
Figure \ref{fig:mounts} indicates where the focii are.
Here we list them, along with examples of the codes for telescopes which use them (in brackets). 
Images of these are
shown in figure 1 of eLBA memo 9.

\begin{itemize}
\item Prime focus (PKS/MED). Focus at the site of the secondary mirror
  (sub-reflector).
\item Cassegrain focus (VLBA/most). Focus after the secondary
  (hyperboloid) mirror.
\item Gregorian focus (EFF). Prime Focus before the secondary
  (ellipsoid) mirror, secondary focus after the secondary mirror.
\item Folded Cassegrain focus (CED). Focus bolted to the elevation axis,
after the tertiary mirror. 
\item Reduced Nasmyth focus (JCMT). Focus on the elevation axis, but
bolted to the azimuth floor, after the tertiary mirror.
\item Full Nasmyth focus (PV/YEB40). Focus bolted to the azimuth axis
floor, after the forth mirror.
\item Beam Wave Guide focus (WARK/YAMA). Focus bolted to the floor of the antenna support structure, after the fifth  (or more) mirror.
\end{itemize}
%

Fixed mirrors after the forth, in a Nasmyth system, swap the mount type
between the equivalent of a reduced and full Nasmyth, so that any
system with an odd number of mirrors is a `reduced' and an even number
of mirrors is equivalent to a `full' Nasmyth system in our context. In
a similar fashion the expression for a folded Cassegrain and Prime
focus solutions are essentially the same, being either one or three
reflections from the on-sky view.
The difference between a Cassegrain and a Gregorian image is a
rotation (of 180$^o$), so these types are degenerate after the absolute
position angle calibration.
Furthermore at the correlator, the `on sky' left or right is combined
with the same to produce the parallel hand output. This is effectively
the same as adding another mirror to the optical chain for those
optics with an odd number of mirrors. For all expected cases,
therefore, we will be correlating either Cassegrain, Nasmyth or Beam WaveGuide foci, combined with the drive type of the antenna. 
Nasmyth is Right or Left handed, depending on the orientation of M3. Likewise M5 can be Right or Left handed, but in all cases we have seen they are R where M3 is L, and vice versa.

\subsection{Drive configuration}

There are three drive configurations used in VLBI. This also effects the
rotation of the feeds as seen from the sky.

\begin{itemize}
\item HA-DEC (WST), for Hour Angle -- Declination. Also called
equatorial. It produces no change of feed angle as the source passes
across the sky. As HA-DEC mounts require an asymmetrical structural
design they are unsuited to the support of very heavy
structures. Their advantage is that motion is required in only
one axis to track an object as Earth rotates.
\item ALT-AZ (VLBA/most), for Altitude -- Azimuth. Also called
Az-El. It rotates the telescope pointing by the parallactic angle as
the source passes across the sky. It is the most common variety of
Radio telescope mount. It is symmetric so can support very large
antennae, but at the zenith small angular changes on the sky can
require very large angular changes in the azimuth angle. This effect
is call the `keyhole'.
\item EW (HOB), for East -- West. It rotates the telescope by the
co-parallactic angle as the source passes across the sky. An EW mount
places the focus very high, therefore it is prone to flexing and
poor pointing. Their advantage is that they can track across the sky at
high speeds and without gaps (i.e. it does not have a `keyhole' at the
zenith). They are primarily used for tracking Low Earth Orbit
satellites.
\end{itemize}

The parallactic angle, sometimes called the position angle, of an
object is the angle between the celestial pole (north or south
depending on location), the object and the zenith (the point in the
sky directly overhead). It is not to be confused for the same named
angle which is also called the convergence angle. This is the
difference in the angular direction to an object at a distance, from
two points of view, i.e. the parallax. The expression for the
parallactic angle as we use it is;

\[ 
\chi_p = {\rm arctan}[\frac{{\rm sin}(\Theta)\ {\rm cos}(l)}
                    {{\rm cos}(\delta){\rm sin}(l) -
		      {\rm cos}(\Theta){\rm cos}(l){\rm sin}(\delta)}]
\]

The co-parallactic angle is;

\[ 
\chi_c = {\rm arctan}[\frac{{\rm cos}(\Theta)}
                    {{\rm sin}(\delta){\rm sin}(\Theta)}]
\]

The Nasmyth angle is almost the same as the parallactic angle, but
with the relative rotation of the forth mirror included;

\[ 
\chi_n = \chi_p \pm E
\]

The BWG angle is almost the same as the Nasmyth angle, but
with the relative rotation of the fifth mirror included;

\[ 
\chi_b = \chi_n \mp A
\]

where $\Theta$ is the hour angle, and $\delta$ the declination, of the
source. $l$ is the latitude of the telescope, $E$ is the
elevation and $A$ is the azimuth. 
As all BWG bring the focus back into the pedistal, the sign of $E$ and $A$ are opposite; in principle this does not need to be the case. 

Each mirror adds a swap of polarisation from LHC to RHC. Therefore, in
principle, the number of reflections before the horn is very important
in any optical system, as each mirror also swaps the feed angle
polarisation vector on the sky.
However when the detected LHC or RHC polarisations are relabelled to
represent the on-sky value, an effective mirror is added to produce
the parallel or cross hand outputs. This means that only Cassegrain, Nasmyth or BWG models are required.

\begin{footnotesize}
\begin{table}[htb]
\begin{center}
\begin{tabular}{lcr}
&&{Mount Number}\\
Focus and Drive&Label&AIPS\\
\hline
Cassegrain and Alt-Az& ALAZ &0\\
Any and Equatorial & EQUA &1\\
Any and Orbiting& ORBI &2\\
\multicolumn{3}{c}{New mount types}\\
Prime focus and EW  & EW-\,- &3\\
Right hand Naysmyth and Alt-Az& NS-R &4\\
Left hand Naysmyth and Alt-Az& NS-L &5\\
Right hand BWG and Alt-Az& BW-R &6\\
Left hand BWG and Alt-Az& BW-L &7\\
\hline
\end{tabular}
\caption{Mount Types}
\end{center}
\end{table}
\end{footnotesize}

\section{The Beam Wave Guide configuration}
\begin{figure}[h]
    \centering
    \includegraphics[width=0.7\textwidth]{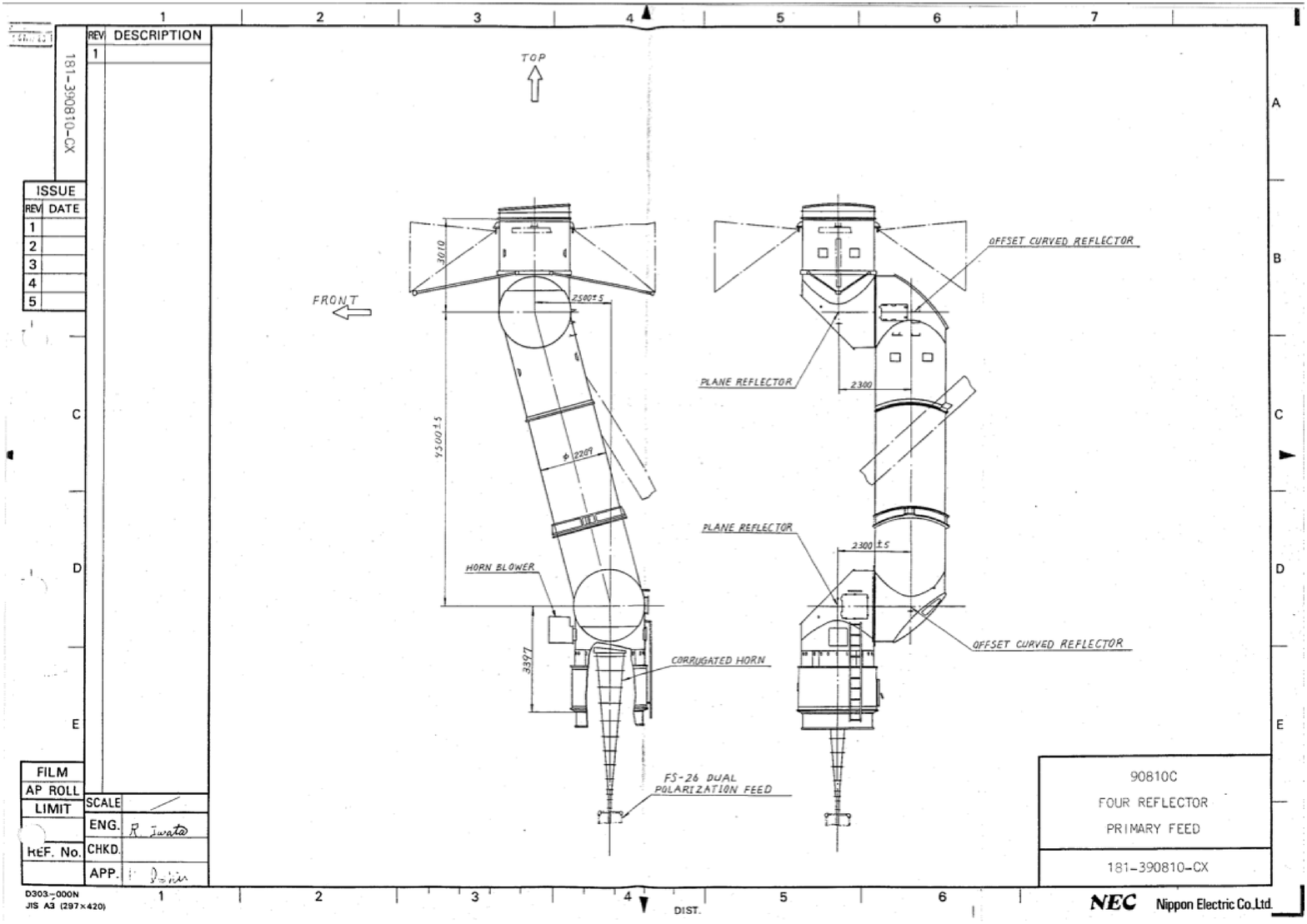}
    \caption{Schematic of the Warkworth 30-m radio telescope optics BWG optics.}
    \label{fig:pf1}
\end{figure}
The feed in a beamed wave guide system sits on the floor on the pedestal, thus avoiding movement and related issues. The image of the sky has to pass over two rotating mirrors, here assumed to be M3 and M5. Other static mirrors can and often are introduced, but the only fact that is important is the rotational behaviour of the mirrors that move.
I.e. for a ``folded cassagrain'' (only Ceduna to our knowledge) the M3 rotates with the elevation axis, so there is no extra term introduced and this behaves as a Cassegrain (if the mount in Alt-Az).
Therefore the difference between the instrumental
polarisation angle of a BWG mount and the parallactic angle is plus or minus the elevation angle and minus or plus the azimuthal angle. 
This assumes that the BWG directs the image of the sky back to the central axis; if it does not the sign of the Azimuth angle changes. 
\begin{figure}
    \centering
    \includegraphics[width=0.7\textwidth]{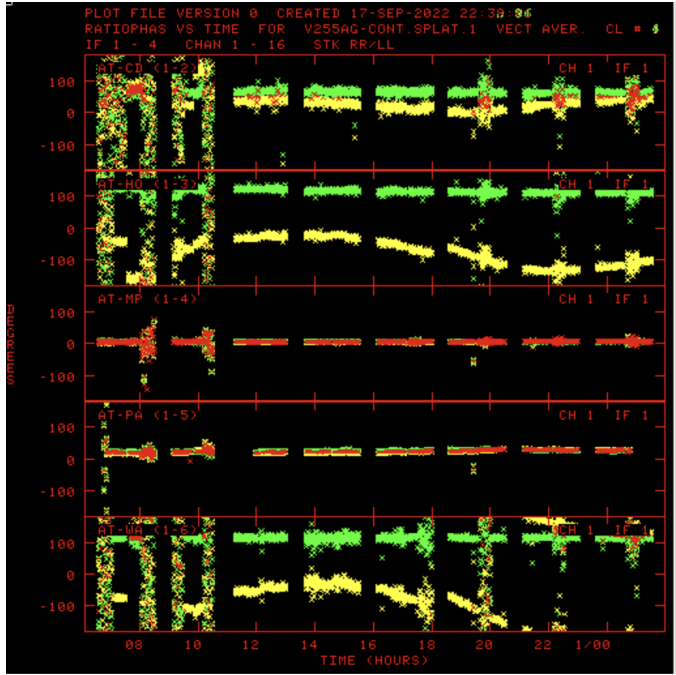}
    \caption{The impact of the R/L phase difference for V255AG, for a number of calibrators, before (Yellow) and after (Green) the application of the mount feed rotation correction, with CLCOR. The antenna mount types are AT, CD, MP, PA all Alt-Az, Hobart EW-mount and Warkworth right handed Beam Wave Guide.}
    \label{fig:step1}
\end{figure}
\begin{figure}
    \centering
    \includegraphics[width=0.7\textwidth]{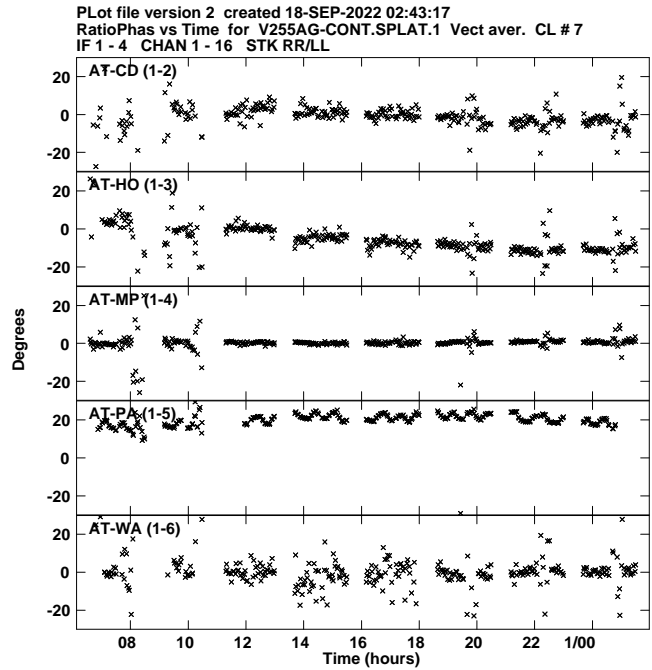}
    \caption{Zoom in on the R/L phase difference for V255AG, after removing a constant R/L phase offset using one scan. Imperfections in the R/L calibration are noticeable. Parkes has an offset (so requires using a different calibration scan) and a ripple, whilst Ceduna and Hobart show slow wandering. This maybe due to poor polarisation purity.
    \label{fig:step2}}
\end{figure}

\pagebreak

\appendix
\section{Antenna Table Example}

This is the table from V255AG after correction using AIPS 31DEC22 with
update UPD20221017.064133.

\begin{quote}
File=V255AG-CONT .SPLAT .   1     An.ver=   1     Vol= 1     User=  255

AT       BX= -4751640.5462 BY=  2791700.2552 BZ= -3200490.1824 Mount=ALAZ  

CD       BX= -3753443.6701 BY=  3912709.8121 BZ= -3348066.5042 Mount=ALAZ  
 
HO       BX= -3950237.6040 BY=  2522347.7251 BZ= -4311561.6218 Mount=EW$--$

MP       BX= -4682769.8739 BY=  2802618.8436 BZ= -3291758.3473 Mount=ALAZ  

PA       BX= -4554232.7144 BY=  2816758.8501 BZ= -3454034.7194 Mount=ALAZ  
 
WA       BX= -5115425.85 BY=   477880.2434 BZ= -3767042.0 Mount=BW$-$R
\end{quote}

Running CLCOR on an array including Hobart (co-parallactic) and Warkworth (BWG right handed) you get:
\begin{quote}
CLCOR1: Task CLCOR  (release of 31DEC22) begins

CLCOR1: Copied CL file from vol/cno/vers  1    1   1 to  1    1   9

CLCOR1: CL version input   1 output   9

CLCOR1: Using CO-PARALACTIC ANGLES for antenna   3

CLCOR1: BeamWaveGuide formula for antenna   6

CLCOR1:   3209 Records modified

CLCOR1: Appears to have ended successfully

\end{quote}

\section{Recommendations for calibration of polarisation observations}

Many references have covered this, but here I wish to clearly layout
the important steps in AIPS and the consequences of the tasks.

\begin{itemize}
\item {\bf TABED} options: INE 'AN'; OPTY 'repl'; APARM 5 0 0 4 4 {\bf
  6} and KEYV {\bf 6} 0\\ Replaces the MNTSTA type of antenna 6 (Wa) with
  value 6 (for a right handed BWG, as is the case for Warkworth).\\
\item {\bf CLCOR} options: CLCORP 1 0 and OPC 'pang'\\
Calculates the phase correction for the listed mount types.
\item {\bf FRING} options: CALS 'prime\_cal'; APARM(3)=0 and APARM(5)=0\\
Finds the independent delays and combined rates and phases for the
calibrator. Now the RR/LL phase will be constant.  
\item {\bf CALIB} options: CALS 'prime\_cal'; SOLMODE 'P'; APARM(3)=0 and
APARM(5)=0\\ 
Finds the independent rates and phases for the calibrator. Now the
RR, LL and RR/LL phases will be zero. One may wish to use only one
scan on the prime calibrator. 
\item {\bf VLBACPOL} this procedure finds the delays between left and
  right hand for the reference antenna. If PCAL (or something similar)
  is used the delay between left and right should be zero.
\item {\bf FRING} options: CALS 'target','calib'; APARM(3)=1 and APARM(5)=1\\
Finds the rates and combined phases for the target.
\item {\bf CALIB} options: CALS 'calib'; SOLMODE 'A\&P'; APARM(3)=1
and APARM(5)=1\\ Produces a well calibrated version of the target
which can be imaged. 
\item {\bf IMAGR}  Deconvolve this with, either a few model
components, or clean it and then box up the clean components into a
few regions (with CCEDT), and use it for the next stage.
\item {\bf LPCAL} options: CALS 'calib'; in2na 'calib' and in2c 'icln'\\
Does the polarisation calibration on the target, using the cleaned
model.
\item {\bf CALIB} options: CALS 'target'; SOLMODE 'A\&P'; DOPOL 2;
APARM(3)=1 and APARM(5)=1\\ 
Calibrates the target using the polarisation solutions.
\item {\bf IMAGR} Image the source in I, Q and U. The sum of the
  fluxes (total and polarised) should compare to the lower resolution
  (VLA or ATCA) values. The correction $\phi_{{\rm RL}}$ is
  $2\chi_{{\rm true}}-\arctan(\frac{\sum{\rm U}}{\sum{\rm Q}})$, for
  each IF.
\item {\bf CLCOR} options: CLCORP  $\phi_{{\rm RL}}$ stokes 'L' and OPC 'polr'\\
  Rotates the D-terms to match the calibration value of $\chi_{{\rm
      true}}$.
\end{itemize}

If this recipe is not followed the solutions for the L and R hand
polarisations are independent. This is not a problem if there is
sufficient signal to noise. 

\section{A list of the changes to classic AIPS 31DEC22}
\label{app:code}

A summary of the files that need to be changed:

\begin{itemize}
\item  PARANG.FOR: all mount types now supported
\item  PARACO.FOR: all mount types now supported (PARACO is a new function almost the same as PARANG)
\item  PRTAN.FOR: Add mount names
\item  CLCOR.FOR: in ANAXIS allow all mount types
\item  DFCOR.FOR: in ANAXIS allow all mount types
\item  SNPLT.FOR: does not use PARANG, so mount types added
\item  DTSIM.FOR: (in ORIENT) does not use PARANG, so mount types added
\item  DTSIM.FOR: (in GETBAS) allow all mount types
\end{itemize}

\end{document}